\documentclass[journal]{IEEEtran}
\usepackage[dvipdfmx]{graphicx,xcolor}
\usepackage{epsfig}
\usepackage[T1]{fontenc}
\usepackage{lmodern}
\usepackage{textcomp}
\usepackage{latexsym}
\usepackage{cite}
\usepackage{bm}
\usepackage{amsmath}
\usepackage{amsfonts}
\newcommand{\vct}[1]{\mbox{\boldmath{$#1$}}}

\newcommand{\jj}        {\mathrm{j}}

\begin{document}
\title{Noncontact Detection of Sleep Apnea Using Radar and Expectation-Maximization Algorithm}
\author{Takato~Koda, Shigeaki Okumura, Hirofumi~Taki~\IEEEmembership{Member,~IEEE}, Satoshi~Hamada, Hironobu Sunadome, Susumu~Sato, Kazuo Chin, and Takuya~Sakamoto~\IEEEmembership{Senior Member,~IEEE}
  \thanks{This work was supported in part by JST under Grants JPMJCE1307 and JPMJMI22J2, in part by SECOM Science and Technology Foundation, and in part by JSPS KAKENHI under Grants 19H02155, 21H03427, and 23H01420.}
\thanks{Takato~Koda and Takuya Sakamoto are with the Graduate School of Engineering, Kyoto University, Kyoto 615-8510, Japan (e-mail: sakamoto.takuya.8n@kyoto-u.ac.jp).}
\thanks{Shigeaki~Okumura and Hirofumi~Taki are with MaRI Company Ltd., Kyoto 600-8815, Japan (e-mail: sokumura@marisleep.co.jp, taki@marisleep.co.jp).}
\thanks{Satoshi~Hamada, Hironobu~Sunadome, Susumu~Sato, and Kazuo Chin are with the Graduate School of Medicine, Kyoto University, Japan. Kazuo~Chin is also with the School of Medicine, Nihon University, Itabashi-ku Tokyo 173-8610, Japan (sh1124@kuhp.kyoto-u.ac.jp, sankichi@kuhp.kyoto-u.ac.jp, ssato@kuhp.kyoto-u.ac.jp, chin.kazuo@nihon-u.ac.jp).}}

\markboth{}%
         {Koda \emph{et al.}: Noncontact Detection of Sleep Apnea Using Radar and Expectation-Maximization Algorithm}
         
\maketitle
\begin{abstract}
  Sleep apnea syndrome requires early diagnosis because this syndrome can lead to a variety of health problems. If sleep apnea events can be detected in a noncontact manner using radar, we can then avoid the discomfort caused by the contact-type sensors that are used in conventional polysomnography. This study proposes a novel radar-based method for accurate detection of sleep apnea events. The proposed method uses the expectation-maximization algorithm to extract the respiratory features that form normal and abnormal breathing patterns, resulting in an adaptive apnea detection capability without any requirement for empirical parameters. We conducted an experimental quantitative evaluation of the proposed method by performing polysomnography and radar measurements simultaneously in five patients with the symptoms of sleep apnea syndrome. Through these experiments, we show that the proposed method can detect the number of apnea and hypopnea events per hour with an error of 4.8 times/hour; this represents an improvement in the accuracy by 1.8 times when compared with the conventional threshold-based method and demonstrates the effectiveness of our proposed method.
\end{abstract}

\IEEEpeerreviewmaketitle

\section{Introduction}
\label{sec:introduction}
The number of adults aged from 30--69 years that display moderate to severe symptoms of sleep apnea syndrome (SAS) is estimated to be approximately 425 million worldwide \cite{osa1}. SAS is known to increase the risk of complications such as hypertension, coronary artery disease, and cerebrovascular disease, and can also lead to work-related errors and traffic accidents because of daytime sleepiness and poor concentration \cite{osa4}. To avoid these risks, early diagnosis and treatment are essential \cite{osa2,osa3}, and the gold standard for SAS testing is polysomnography (PSG) \cite{aasm1}. PSG records various physiological signals, including electroencephalography (EEG), electrocardiography (ECG), and arterial blood oxygen saturation signals throughout the night using contact-type sensors. When compared with the conventional PSG technique, radar-based noncontact SAS monitoring offers the advantage that the discomfort caused to the patient by contact-type sensors can be avoided, thus enabling monitoring of the SAS signs during natural sleep.

Many existing studies of noncontact detection of sleep apnea (SA) are based on machine learning (ML) methods \cite{Kwon1,Anishchenko,Yang,Javaid,Islam,Koda,Snigdha,Fukuyama}, including long short-term memory (LSTM) \cite{Kwon1,Anishchenko,Yang} and support vector machine (SVM) techniques \cite{Islam,Koda,Snigdha}. Although ML-based methods often require sufficient quantities of training data to be available for accurate classification, it is not easy to collect enough data effectively because the radar data depend on the measurement setup and the surrounding environment, and also depend on the positioning and angle of the radar system \cite{Kwon1,Anishchenko}, thus limiting the use of ML in practice.

In contrast, test methods without use of ML have also been reported \cite{Xiong1,Xiong2,Baboli,Du,Kagawa1,Kagawa2,Qi1,Qi2,Khushaba,Fedele,Kang}, and some of these methods are based on thresholding in terms of the displacement amplitude. For example, in a study by Kagawa {\it et al.} \cite{Kagawa1,Kagawa2}, a threshold was set up using the amplitude of the displacement caused by normal respiration. Another example is a study by Kang {\it et al.} \cite{Kang} that proposed a method for use of adaptive thresholding in a constant false alarm rate (CFAR) framework. These methods require prior information about the normal respiration of the target patient, which limits the practicality of the methods.

Unlike conventional ML-based or threshold-based methods, this study introduces a new approach that uses the statistical distribution of respiratory displacement. The proposed method uses the expectation-maximization (EM) algorithm to estimate parameters for stochastic models of respiratory displacement and distinguish between normal breathing and apnea to detect apnea without using numerous training datasets or threshold values. In this study, we conducted experiments on five patients who had been hospitalized for SA testing using PSG, and evaluated the accuracy of apnea event detection using the proposed method. Through these experiments, we demonstrate that the proposed method can detect apnea events with higher accuracy than the conventional threshold-based method. 

\section{Sleep Apnea Detection Using Array Radar}
\subsection{Millimeter-Wave Array Radar}

In this study, we apply a millimeter-wave radar system that uses the frequency-modulated continuous-wave (FMCW) method with a center frequency of 79 GHz, a center wavelength of $\lambda=3.8$ mm, range resolution of 43 mm, transmit power of 9 dBm, an equivalent isotropic radiated power of 20 dBm, and a slow-time sampling frequency of 10 Hz.
The antennas of this radar system comprise a multiple-input multiple-output (MIMO) array that contains three transmitting elements and four receiving elements, corresponding to a 12-element virtual array; the transmitting and receiving arrays are linear arrays that are equally spaced at intervals of $2\lambda$ (7.6 mm) and $\lambda/2$ (1.9 mm), respectively. The radiation pattern for each element is defined as $\pm 4^\circ $ and $\pm 35^\circ $ in the E- and H-planes, respectively. The three transmitting elements radiate signals one-by-one in a time-division multiplexing manner.

\subsection{Radar Imaging and Extraction of Respiratory Displacements}
\label{subsec:2-2}
By applying a Fourier transform to the FMCW radar data with regard to the fast time, we obtain a signal $s'_k(t,r)$, where $t$ is the slow time, $r$ is the range, and $k (=0,1,\cdots,K-1)$ is the element number of the virtual array, with the number of virtual elements being denoted by $K=12$. Note that the phase of the signal $s'_k(t,r)$ was assumed to be calibrated in advance.

By assuming that the the $k$th virtual array element is located at $(x,y)=(x_k,0)$ in the $xy$ plane, where $x_k=k\lambda/2$, the signal vector $\vct{s}(t,r)$ can then be obtained as 
\begin{equation}
  \vct{s}(t,r)=[s_0(t,r),s_1(t,r),\cdots,s_{K-1}(t,r)]^\mathrm{T},
\end{equation}
where $s_k(t,r)=c_ks'_k(t,r)$, $c_k$ is a Taylor window coefficient and the superscript $\mathrm{T}$ denotes a matrix/vector transpose operator. 
Using the beamformer weight vector $\vct{w}(\theta)=[w_0,w_1,\cdots,w_{K-1}]^\mathrm{T}$ along with $w_k(\theta)=\mathrm{e}^{-\jj (2\pi x_k/\lambda)\sin\theta}=\mathrm{e}^{-\jj\pi k\sin\theta}$ $(k=0,1,\cdots,K-1)$, a complex-valued radar image $I_0(t,r,\theta)=\vct{w}^\mathrm{H}(\theta)\vct{s}(t,r)$ is then generated, where the superscript $\mathrm{H}$ denotes the complex conjugate transpose operator.

We then suppress the static clutter components as follows: 
\begin{align}
    I_\mathrm{c}(t,r,\theta)=I_0(t,r,\theta)-\frac{1}{T_\mathrm{c}} \int_{t-T_\mathrm{c}}^{t}  I_0(\tau,r,\theta)\,\mathrm{d}\tau,
\end{align}
where $I_\mathrm{c}(t,r,\theta)$ is a clutter-free complex-valued radar image and $T_\mathrm{c}$ will be defined later in the paper.

If the signal received contains only a single dominant echo from the target human body, which is located at $(r_0,\theta_0)$, then the body displacement $d_0(t)$ is obtained as
\begin{align}
    d_0(t)=\frac{\lambda}{4\pi}\angle I_\mathrm{c}(t,r_0,\theta_0).
    \label{eq:disp0}
\end{align}
The respiratory displacement $d(t)$ is estimated using a band-pass filter to suppress the body motion and the random noise components. The cutoff frequencies of the filter are set at 0.17 Hz and 1.8 Hz, corresponding to 0.55 s and 6 s in the time domain, respectively.

\subsection{Limitations of the Conventional Apnea Detection Method}
\label{subsec:2-3}
According to the American Academy of Sleep Medicine (AASM) scoring criteria for SAS, ventilation is reduced by 30\% when compared with normal breathing during hypopnea and by more than 90\% during apnea \cite{scoringmanual,osa1,aasm_standard1, aasm_standard1,aasm_standard2,Troester}, and any such events that last longer than 10 s are counted as hypopnea or apnea, respectively \cite{aasm_standard1,aasm_standard2}. Therefore, the body motion caused by respiration is reduced during hypopnea and apnea events, and many conventional methods detect this reduced body motion. One such example is the amplitude baseline method (ABM) \cite{Kagawa1,Kagawa2}, which is referred to as the conventional method for the purposes of performance comparison in this study.

The displacement amplitude $\bar{d}(t)$ is given by
\begin{align}
    \bar{d}(t)=\sqrt{\frac{1}{T_\mathrm{a}}\int_{t-T_\mathrm{a}/2}^{t+T_\mathrm{a}/2} |d(\tau)|^2 \,\mathrm{d}\tau },
    \label{eq:amp}
\end{align}
where we set $T_\mathrm{a}=5$ s because the typical respiratory interval is usually shorter than 5 s. The ABM first estimates the body displacement amplitude $\bar{d}_\mathrm{b}$ during normal breathing to act as a baseline, and then sets a threshold $\beta \bar{d}_\mathrm{b}$ to detect reduced displacement. Based on consideration of the AASM criteria, the threshold value is set at either 70\% or 50\% ($\beta=0.7$ or $0.5$) of the amplitude for normal breathing $\bar{d}_\mathrm{b}$. The ABM detects the apnea and hypopnea event period $[t_1, t_2]$, where $\bar{d}(t)<\beta\bar{d}_\mathrm{b}$ holds for any value of $t$ that satisfies the relation $t_1\leq t \leq t_2$. In addition, the condition $t_\mathrm{2}-t_\mathrm{1}\geq T_\mathrm{m} = 10$ s must be satisfied according to the AASM criteria.

The limitation of the ABM lies in the fact that normal breathing must first be detected as a baseline, which is not always easy in practice. In addition, the accuracy of the ABM decreases when the patient's position and posture change, which is the reason why a new method is proposed in the next section.

\section{Proposed Apnea Detection Method}
\subsection{Respiratory Features Extraction with EM Algorithm}
\label{subsec:3-1}
We propose a method that uses the EM algorithm for apnea detection. In this method, we regard the respiratory displacement amplitude $\bar{d}$ as a random variable and assume that $\bar{d}$ follows a Gaussian distribution with a mean $\mu_1$ and a variance $\varSigma_1$ during apnea and hypopnea events; additionally, we assume that $\bar{d}$ also follows a Gaussian distribution with a mean $\mu_2$ and a variance $\varSigma_2$ during normal breathing periods. The respiratory displacement amplitude, including both the apnea and normal breathing, can then be expressed as a Gaussian mixture model.

First, we define the mixing ratios $\pi_1$ and $\pi_2$ ($\pi_1+\pi_2=1$), which represent the ratios of the time lengths of apnea and normal breathing. Using the vectors $\vct{\mu}=[\mu_1,\mu_2]^\mathrm{T}$, $\vct{\varSigma}=[\varSigma_1,\varSigma_2]^\mathrm{T}$, and $\vct{\pi}=[\pi_1,\pi_2]^\mathrm{T}$, the probability density function $G(\bar{d}|\vct{\pi},\vct{\mu},\vct{\varSigma})$ of $\bar{d}$ is given by:
\begin{align}
    G(\bar{d}|\vct{\pi},\vct{\mu},\vct{\varSigma}) = \sum_{k = 1}^{K_\mathrm{e}}  \pi_k\mathcal{N} (\bar{d}|\mu_k,\varSigma_k),
    \label{eq:gauss}
\end{align}
where $\mathcal{N}(\cdot|\mu,\varSigma)$ is the probability density function of a normal distribution with a mean $\mu$ and a variance $\varSigma$, and the number of mixture components $K_\mathrm{e}$ is set to be $K_\mathrm{e}=2$.

If we assume that $\bar{d}$ is observed $M$ times (i.e., $\bar{d}_i$ $(i=1,2,\cdots,M)$), then the log-likelihood function $L(\bar{d}_i|\vct{\pi},\vct{\mu},\vct{\varSigma})$ is given as follows:
\begin{align}
    L(\bar{d}_i|\vct{\pi},\vct{\mu},\vct{\varSigma})=&\log\prod_{i=1}^{M} G(\bar{d}_i|\vct{\pi},\vct{\mu},\vct{\varSigma})\nonumber\\
    =&\sum_{i = 1}^{M} \log\left[\sum_{k = 1}^{K_\mathrm{e}} \pi_k \mathcal{N}(\bar{d}_i|\mu_k,\varSigma_k)\right].
    \label{eq:logL}
\end{align}
To determine the parameters $\vct{\mu}$ and $\vct{\pi}$, we use the EM algorithm, which consists of an expectation (E) step that calculates the posterior probability over the hidden variable $\gamma_{k,i}$ using the current parameters $\pi_k$, $\mu_k$, and $\varSigma_k$, and a maximization (M) step that estimates the parameters $\pi_k$, $\mu_k$ and $\varSigma_k$ using the current value of $\gamma_{k,i}$. Here, $z_{i,k}$ denotes the probability that the $i$th data sample $\bar{d}_i$ is generated by the $k$th Gaussian distribution $\pi_k\mathcal{N} (\bar{d}|\mu_k,\varSigma_k)$. $\gamma_{k,i}$ is estimated as follows:
\begin{align}
    \gamma_{k,i} = \frac{\pi_k\mathcal{N} (X_i|\mu_k,\varSigma_k)}{\sum_{k = 1}^{K_\mathrm{e}} \pi_k \mathcal{N}(X_i|\mu_k,\varSigma_k)}.
    \label{eq:burden}
\end{align}
The optimal parameters $\pi_k^*$, $\mu_k^*$, and $\varSigma_k^*$ are given by:
\begin{align}
    \pi_k^* =& \frac{M_k}{M}\\
    \mu_k^* =& \frac{1}{M_k}\sum_{i = 1}^{M}\gamma_{k,i}X_i\\
    \varSigma_k^* =& \frac{1}{M_k}\sum_{i = 1}^{M}\gamma_{k,i}(X_i-\mu_k^*)^2,
\end{align}
where $M_k = \sum_{i = 1}^{M} \gamma_{k,i} $.

\subsection{Proposed Apnea Detection Method}
\label{subsec:3-2}
\begin{figure}[t]
    \centering
    \includegraphics[width = 0.48\textwidth]{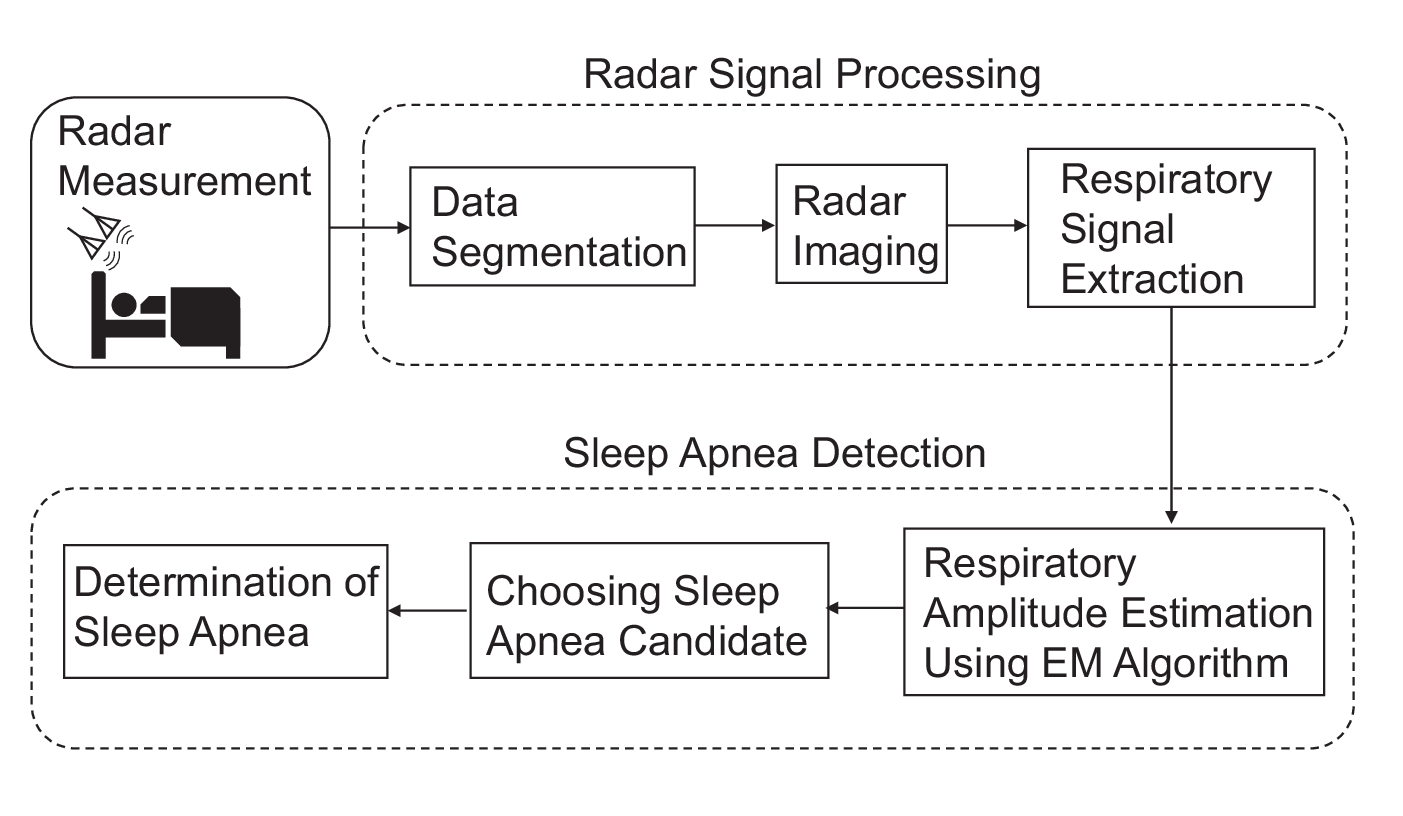}
    \caption{Overview of the proposed apnea detection method.}
    \label{fig:Flow}
\end{figure}

We propose a radar-based system for apnea detection, as illustrated in Fig. \ref{fig:Flow}. The proposed system consists of a radar signal processing block and a sleep apnea (SA) detection block.

First, the overnight radar signal data are divided into epochs, with each having a period of $T_\mathrm{ep}=60$ s, where the adjacent epochs overlap by 30 s following the convention of the PSG test procedures. Next, the radar image $I_\mathrm{p}(r,\theta)$ is generated as follows:
\begin{align}
    I_\mathrm{p}(r,\theta) = \frac{1}{T_\mathrm{ep}}\int_{0}^{T_\mathrm{ep}} |I_\mathrm{c}(t,r,\theta)|^2 \,\mathrm{d}t. 
\end{align}

From the radar image $I_\mathrm{p}(r,\theta)$, local maxima with intensities greater than a specific threshold are extracted, and their range and azimuth characteristics are obtained from $(r_m,\theta_m)$ $(m = 1,2,\cdots,M)$, which correspond to multiple scattering centers on the human body. From the results for the positions $(r_m,\theta_m)$, we obtain a respiratory displacement $d_m(t)$ and an amplitude $\bar{d}_m(t)$ using Eq. (\ref{eq:disp0})--(\ref{eq:amp}). The proposed method uses the displacement waveform not only at a single scattering center but at multiple scattering centers across the human body, because the apnea and hypopnea movements vary depending on the body part involved \cite{aasm_standard2}. By applying the EM algorithm to the displacement amplitude $\bar{d}_m(t)$, the parameters $\pi_{k,m}, \mu_{k,m}, and \varSigma_{k,m}$ $(k=1,2)$ are estimated, where $\mu_{1,m}\leq\mu_{2,m}$. 

Let us define a label $l_m(t)$ as $l_m(t)=1$ if an apnea and hypopnea event is detected from the $m$th displacement amplitude at a time $t$, and let $l_m(t)=0$ otherwise. The proposed method estimates this label as follows: 
\begin{align}
    l_m(t)= 
    \begin{cases}
        1, & \mathrm{if}\;\;\gamma_{1,m}(t)\leq\gamma_{2,m}(t),\;\mathrm{and}\; \displaystyle \frac{\mu_{1,m}}{\mu_{2,m}}\leq\beta;\\
        0, & \mathrm{otherwise.}
    \end{cases}
    \label{eq:l_m}
\end{align}
Here, we set $\beta=0.5$ based on the AASM scoring criteria that were established in 1999 \cite{osa1,aasm_standard1,aasm_standard2}. 
In Eq. (\ref{eq:l_m}), the presence or absence of apnea and hypopnea events is determined by the ratio of $\mu_{1,m}$ to $\mu_{2,m}$.
Finally, we make our decision by using all $l_m(t)$ ($m=1,\cdots,M$) as shown:
\begin{align}
    l(t)= 
    \begin{cases}
        1, & \mathrm{if}\;\; \sum_{m = 1}^{M}  l_m(t)\mu_{2,m}>\sum_{m = 1}^{M}\mu_{2,m}/2  ;\\
        0, & \mathrm{otherwise.}
    \end{cases}
    \label{eq:l}
\end{align}
This approach detects the presence of an apnea and hypopnea event if that apnea and hypopnea event is detected at the majority of the patient's body parts in the radar signals. Because each epoch has a 50\% overlap with its neighboring epoch, the sum of the labels $l(t)$ for each of the apnea events must be either 0, 1, or 2. Therefore, the proposed method only detects the presence of an apnea and hypopnea event when this sum value reaches 2 and the event lasts for more than 10 s.

\newpage
\section{Experimental Performance Evaluation}
\subsection{Overview of the Experimental Setup}
\begin{figure}[t]
    \centering
    \includegraphics[width = 0.25\textwidth]{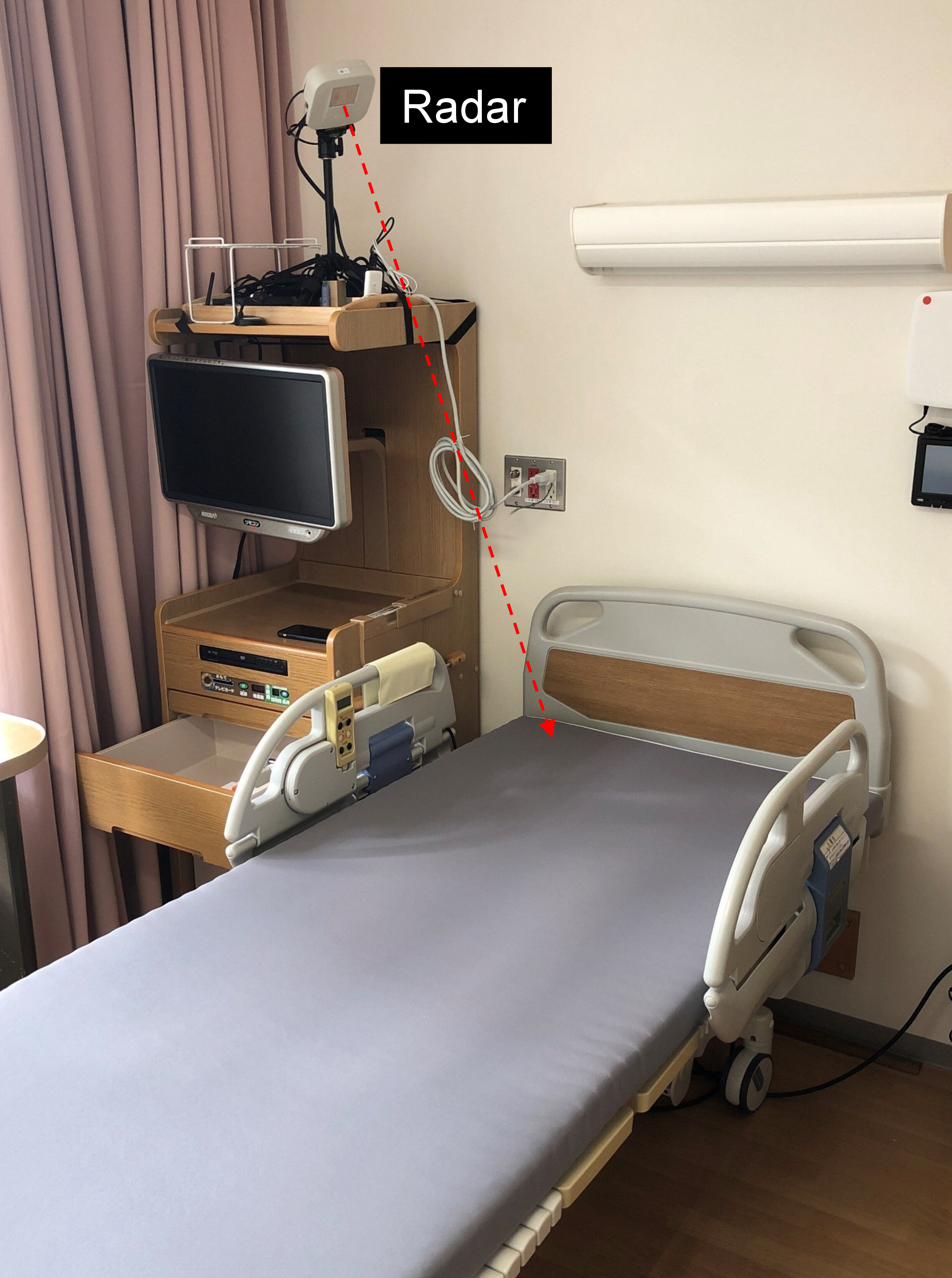}
    \caption{Measurement setup with millimeter-wave array radar system in a hospital room.} 
    \label{fig:hospital}
\end{figure}

To evaluate the performance of the proposed method, we performed radar measurements on patients with SAS symptoms, where the radar and PSG measurements were performed simultaneously, and an evaluation was also performed using conventional medical diagnostic results. The details of the five participating patients with minor to moderate SAS symptoms are presented in Table \ref{tab:Patients}.
Fig.~\ref{fig:hospital} shows the measurement setup with the radar system in a hospital room. Because we are interested in apnea and hypopnea events in this work, we only analyzed the radar data acquired when the patients were asleep. The awakeness of each patient was detected using the PSG data. In the next section, we apply the proposed method to the radar data and compare the results with the PSG data to evaluate the method's performance.

\begin{table}[tb]
    \begin{center}
      \caption{Patient Details}
      \begin{tabular}{|c||c|c|} \hline
        Patient & Radar measurement &Number of\\
        number & time (h) &apnea events\\
        \hline
        1 &  7.0 & 104 \\
        2 &  6.5 & 154 \\
        3 &  5.5 & 68 \\
        4 &  7.0 & 25 \\
        5 &  8.0 & 117 \\
         \hline 
      \end{tabular}
      \label{tab:Patients}
    \end{center}
\end{table}

\subsection{Application of the Proposed Method}
\begin{figure}[t]
    \centering
    \includegraphics[width = 0.4\textwidth]{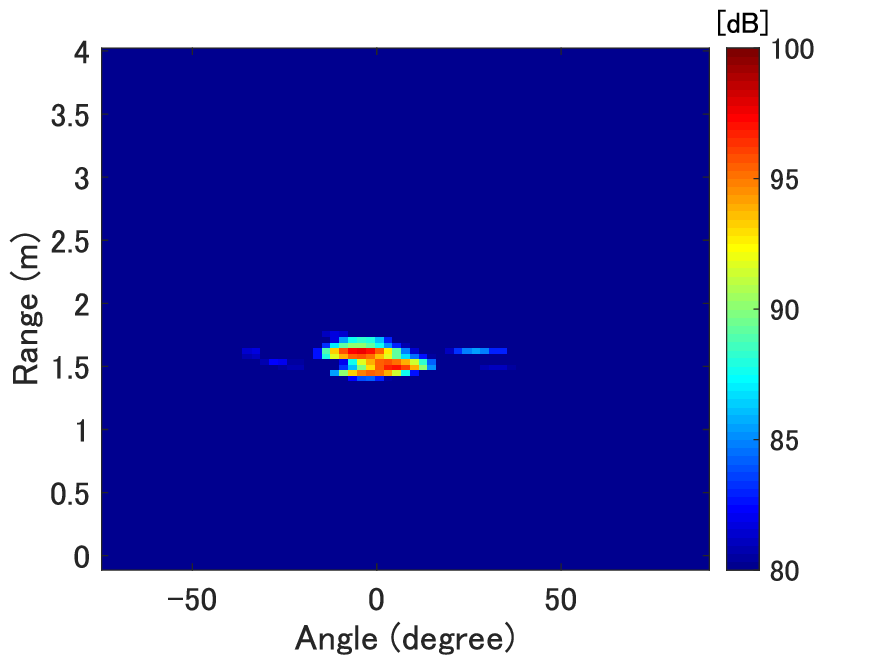}
    \caption{Example of the radar image $I_\mathrm{p}(r,\theta)$.}
    \label{fig:RadarImage}
\end{figure}

In this section, we apply the proposed method to the experimental data to evaluate the method's performance. Fig. \ref{fig:RadarImage} shows a radar image $I_\mathrm{p}(r,\theta)$ that was obtained from data acquired while the patient was asleep, in which we see a set of strong echoes at a range of approximately 1.5 m; this suggests that these echoes arrive from not a single point on the target person's body, but from multiple points. The body displacement waveforms measured at these multiple points were then used to detect apnea and hypopnea events when using the proposed method. 

\begin{figure}[t]
    \begin{center}
        \begin{tabular}{cc}
            \begin{minipage}[t]{0.23\textwidth}
                \centering
                \includegraphics[width = \textwidth]{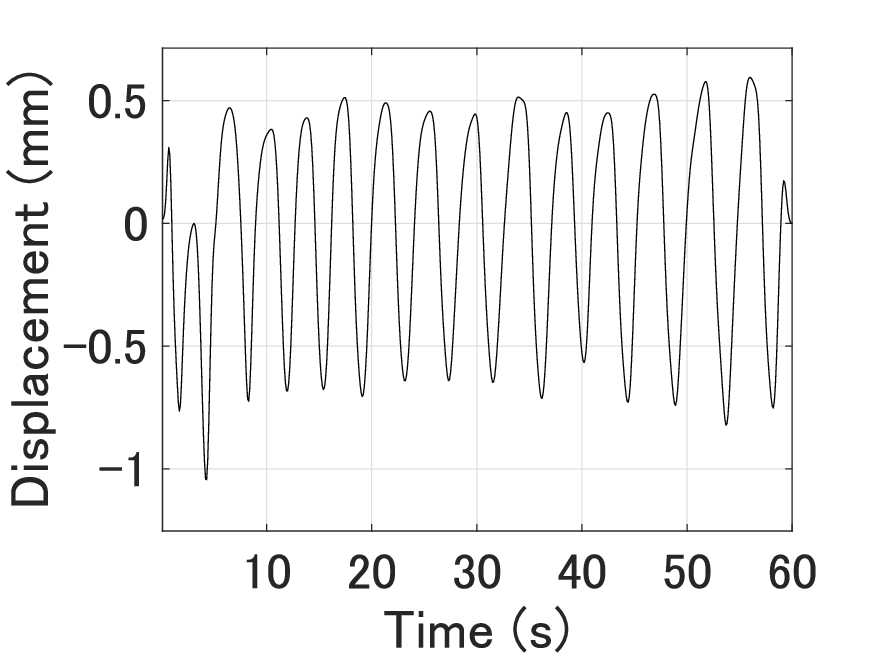}
                \\(a)
            \end{minipage}
            \begin{minipage}[t]{0.23\textwidth}
                \centering
                \includegraphics[width = \textwidth]{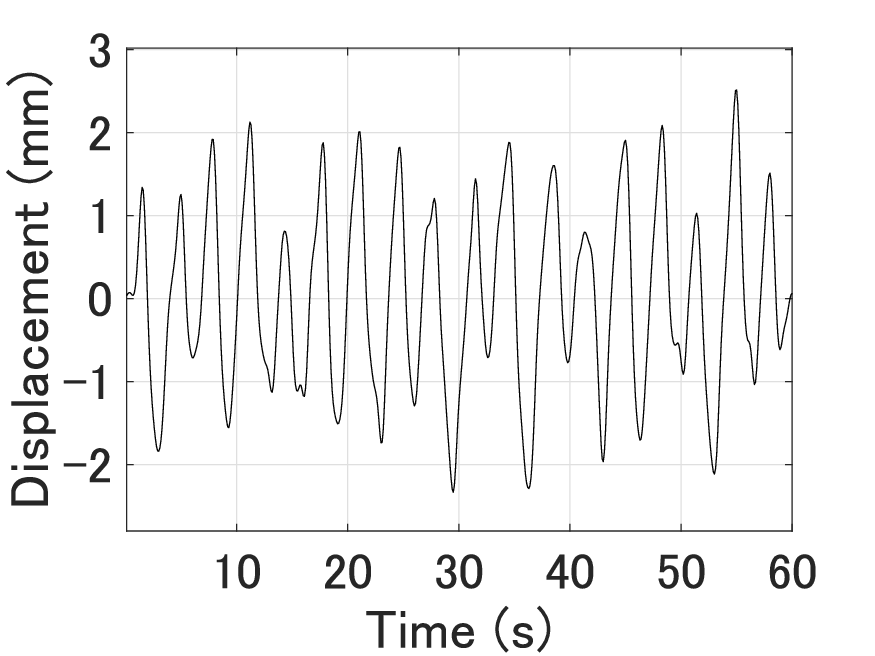}
                \\(b)
            \end{minipage}  
        \end{tabular}
        \\\vspace{2mm}
        \begin{tabular}{cc}
            \begin{minipage}[t]{0.23\textwidth}
                \centering
                \includegraphics[width = \textwidth]{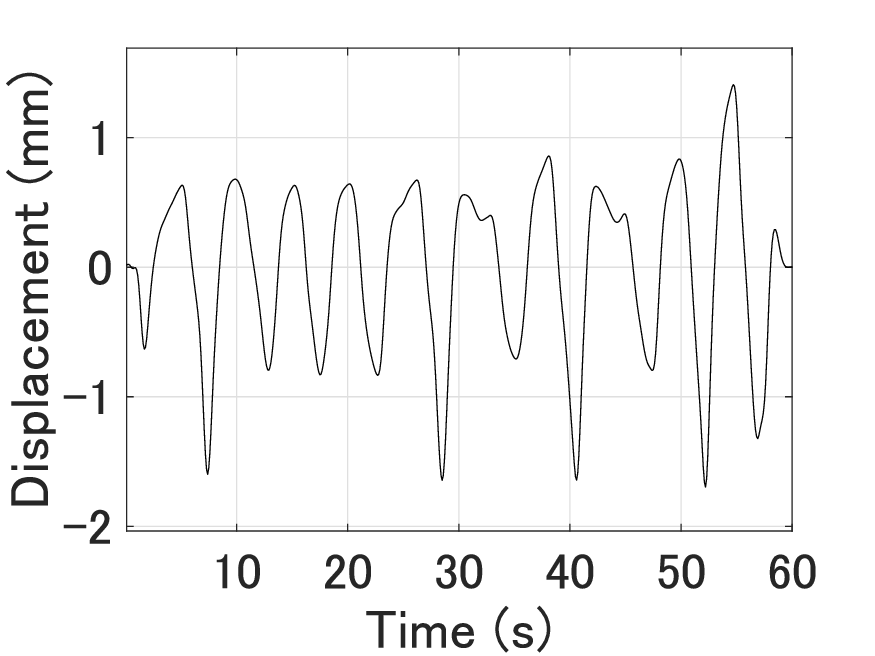}
                \\(c)
            \end{minipage}
            \begin{minipage}[t]{0.23\textwidth}
                \centering
                \includegraphics[width = \textwidth]{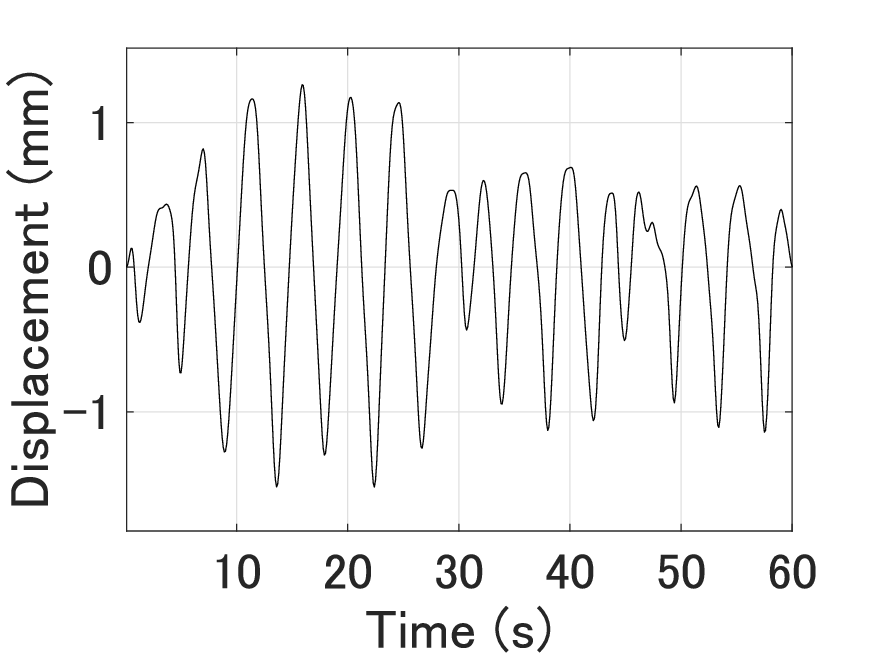}
                \\(d)
            \end{minipage}  
        \end{tabular}
    \end{center}
    \vspace{2mm}
    \caption{Examples of respiratory displacements observed during normal breathing ((a) participant 1, (b) participant 3, (c) participant 4, and (d) participant 5).}
    \label{fig:disp_normal}
\end{figure}

\begin{figure}[t]
    \begin{center}
        \begin{tabular}{cc}
            \begin{minipage}[t]{0.23\textwidth}
                \centering
                \includegraphics[width = \textwidth]{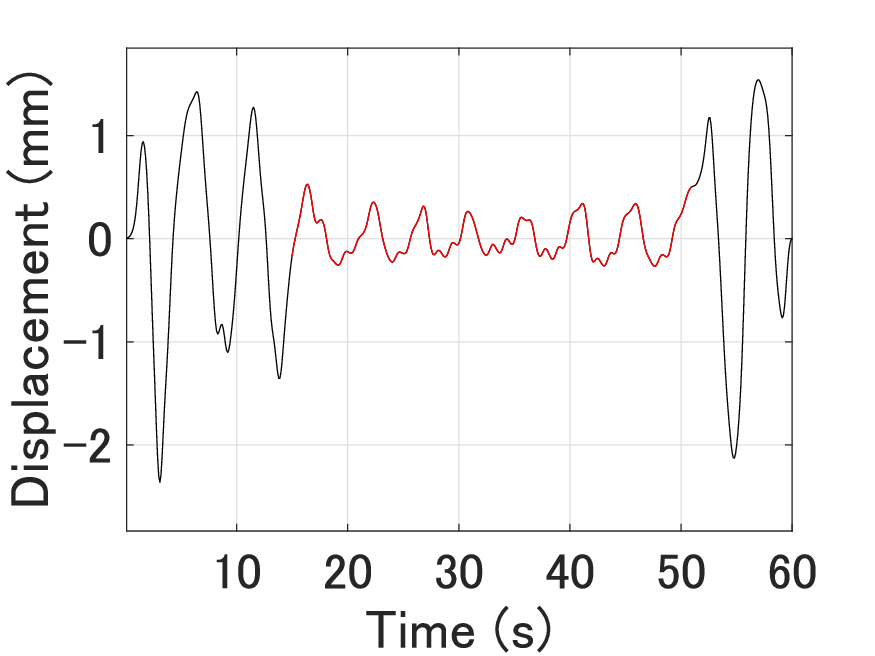}
                \\(a)
            \end{minipage}
            \begin{minipage}[t]{0.23\textwidth}
                \centering
                \includegraphics[width = \textwidth]{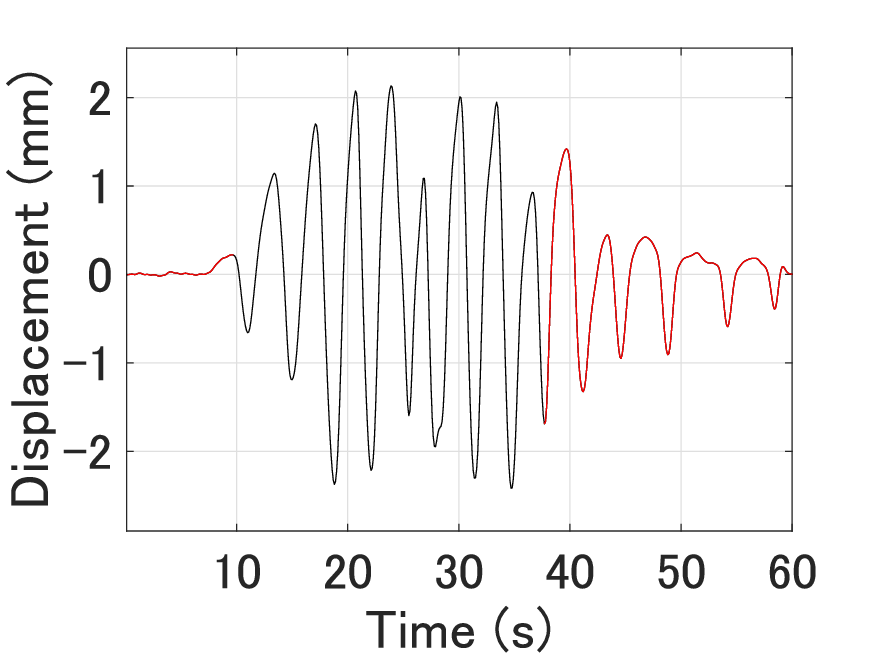}
                \\(b)
            \end{minipage}  
        \end{tabular}
        \\\vspace{2mm}
        \begin{tabular}{cc}
            \begin{minipage}[t]{0.23\textwidth}
                \centering
                \includegraphics[width = \textwidth]{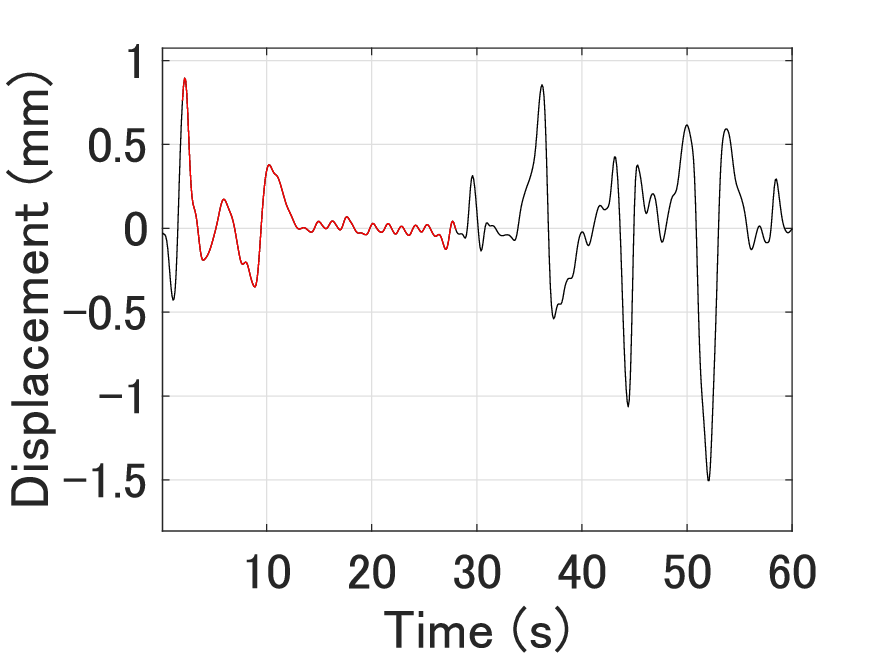}
                \\(c)
            \end{minipage}
            \begin{minipage}[t]{0.23\textwidth}
                \centering
                \includegraphics[width = \textwidth]{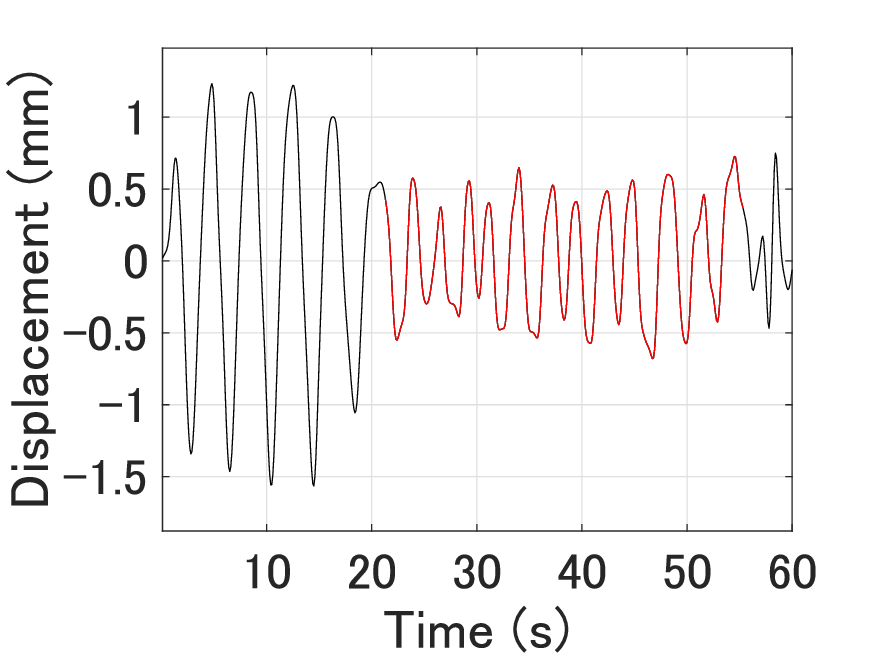}
                \\(d)
            \end{minipage}  
        \end{tabular}
    \end{center}
    \vspace{2mm}
    \caption{Examples of respiratory displacements observed when apnea and hypopnea events occur ((a) participant 1, (b) participant 3, (c) participant 4, and (d) participant 5).}
    \label{fig:disp_apnea}
\end{figure}

Figs. \ref{fig:disp_normal} and \ref{fig:disp_apnea} show examples of the respiratory displacements measured using the radar system during normal breathing and during apnea and hypopnea events, respectively. In Fig. \ref{fig:disp_normal}, the respiratory amplitudes remained almost constant within the 60 s epoch in panels (a)--(d). In contrast, in Fig. \ref{fig:disp_apnea}, the apnea and hypopnea respiratory amplitudes were reduced (red lines) when compared with normal respiration (black lines). Panels (a) and (d) show the obstructive sleep apnea (OSA) signs and panel (b) shows the hypopnea sign. In these three panels, respiratory efforts, i.e., periodic movements similar to those of normal breathing when the amplitude is decreasing, were found. However, no respiratory effort was observed in panel (c), showing the central sleep apnea (CSA) sign. Therefore, apnea and hypopnea events can be determined visually to some extent from the respiratory displacements measured using radar. We note that all participants' apnea and hypopnea may have included central apnea and hypopnea, nontheless, most SAS patients have OSA symptoms, the detection of OSA events is considered important.


\begin{figure}[tb]
    \begin{center}
        \begin{tabular}{cc}
            \begin{minipage}[t]{0.23\textwidth}
                \centering
                \includegraphics[width = \textwidth]{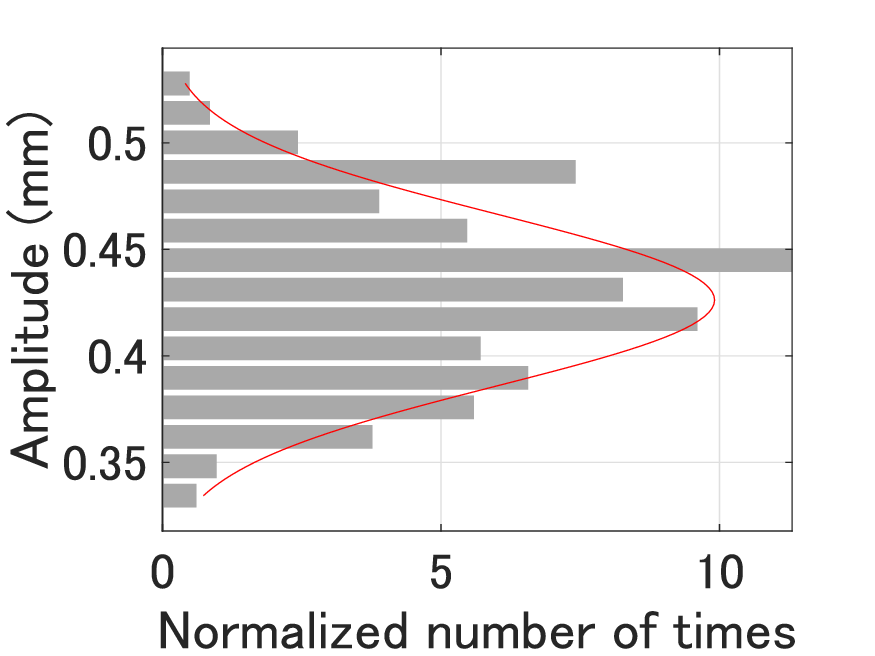}
                \\(a)
            \end{minipage}
            \begin{minipage}[t]{0.23\textwidth}
                \centering
                \includegraphics[width = \textwidth]{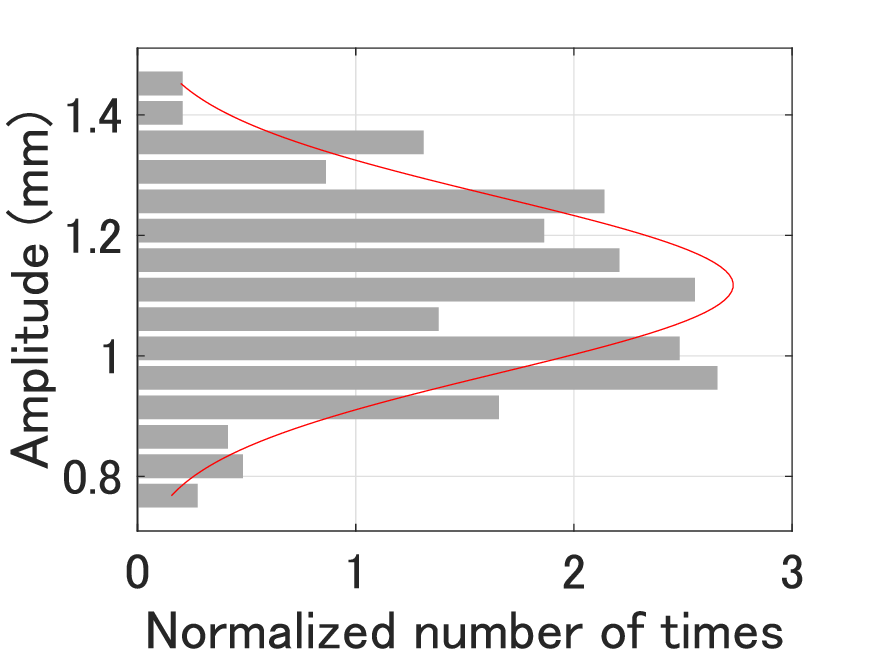}
                \\(b)
            \end{minipage}  
        \end{tabular}
        \\\vspace{2mm}
        \begin{tabular}{cc}
            \begin{minipage}[t]{0.23\textwidth}
                \centering
                \includegraphics[width = \textwidth]{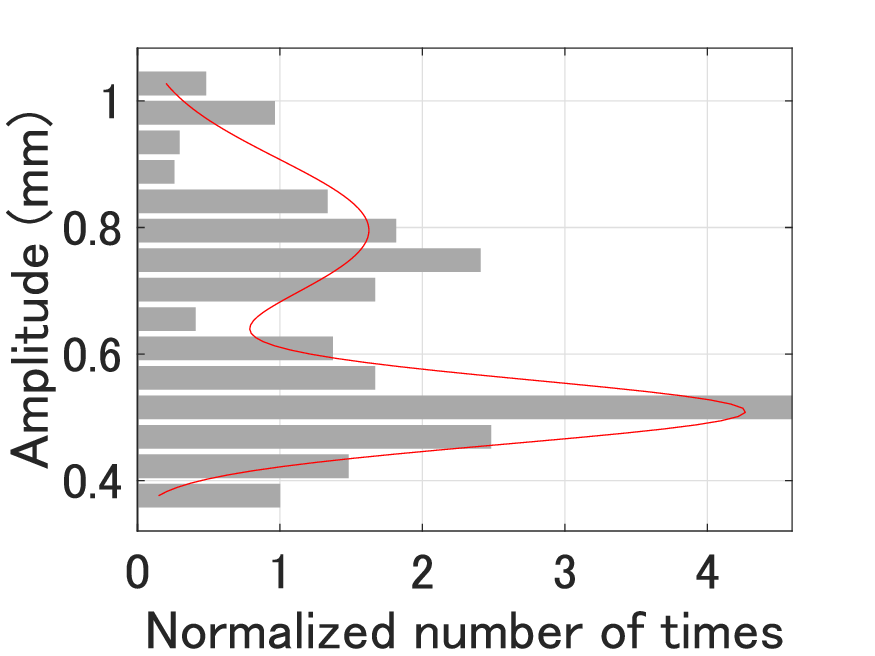}
                \\(c)
            \end{minipage}
            \begin{minipage}[t]{0.23\textwidth}
                \centering
                \includegraphics[width = \textwidth]{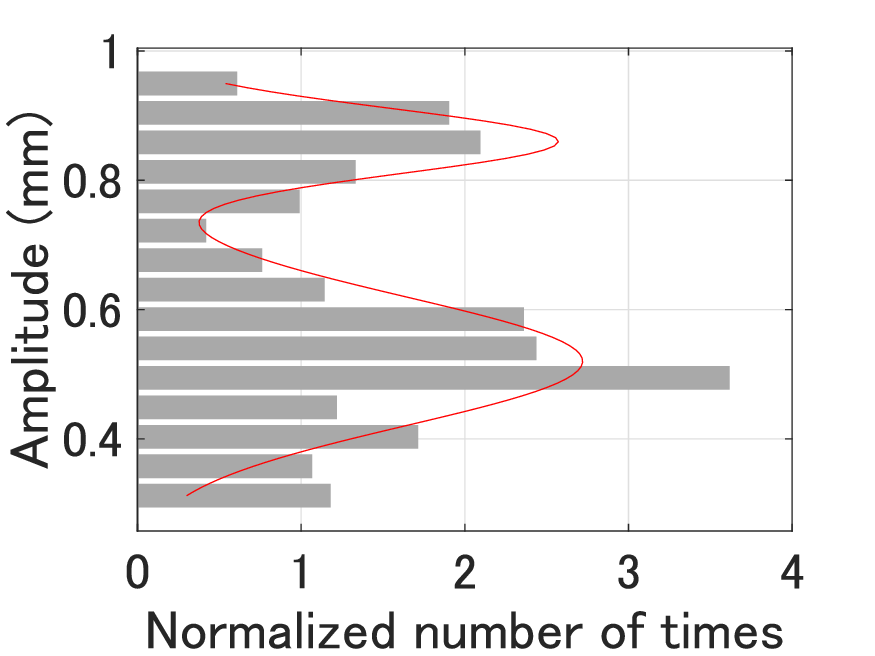}
                \\(d)
            \end{minipage}  
        \end{tabular}
    \end{center}
    \vspace{2mm}
    \caption{Probability density functions estimated using the EM algorithm for respiratory amplitudes during normal breathing ((a) participant 1, (b) participant 3, (c) participant 4, and (d) participant 5).}
    \label{fig:EM_normal}
\end{figure}

Fig. \ref{fig:EM_normal} shows examples of the histograms of the respiratory amplitudes (gray bars) and the corresponding probability density functions of the mixed Gaussian distributions that were estimated using the EM algorithm, where each panel (a)--(d) corresponds to panels (a)--(d) in Fig. \ref{fig:disp_normal}. In panels (a) and (b), only one peak is observed, whereas in panels (c) and (d), we can see two peaks that approximately correspond to $\mu_1 and \mu_2$ $(\mu_1<\mu_2)$. This illustrates that there can be two peaks in these characteristics, even when the breathing is normal.

\begin{figure}[tb]
    \begin{center}
        \begin{tabular}{cc}
            \begin{minipage}[t]{0.23\textwidth}
                \centering
                \includegraphics[width = \textwidth]{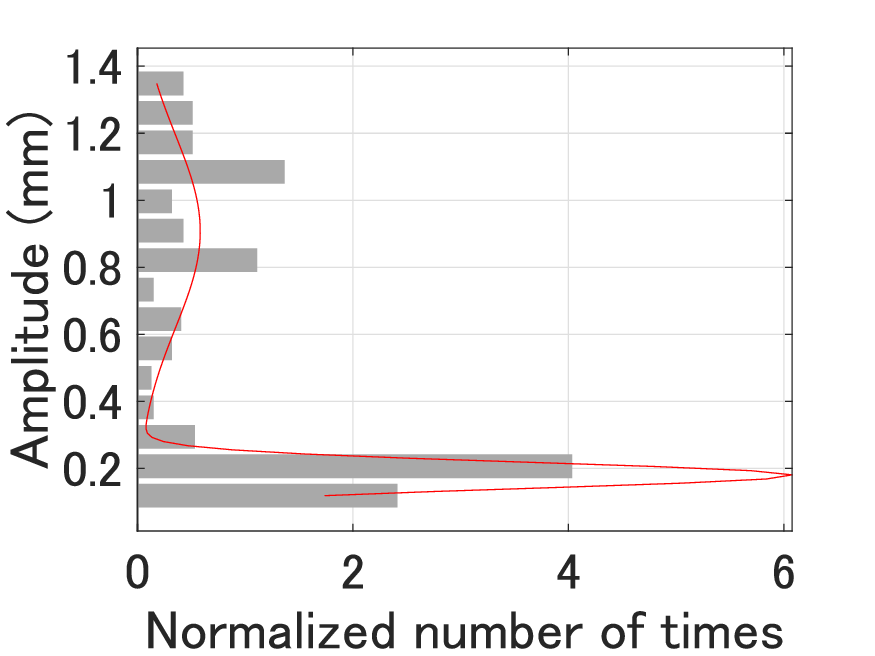}
                \\(a)
            \end{minipage}
            \begin{minipage}[t]{0.23\textwidth}
                \centering
                \includegraphics[width = \textwidth]{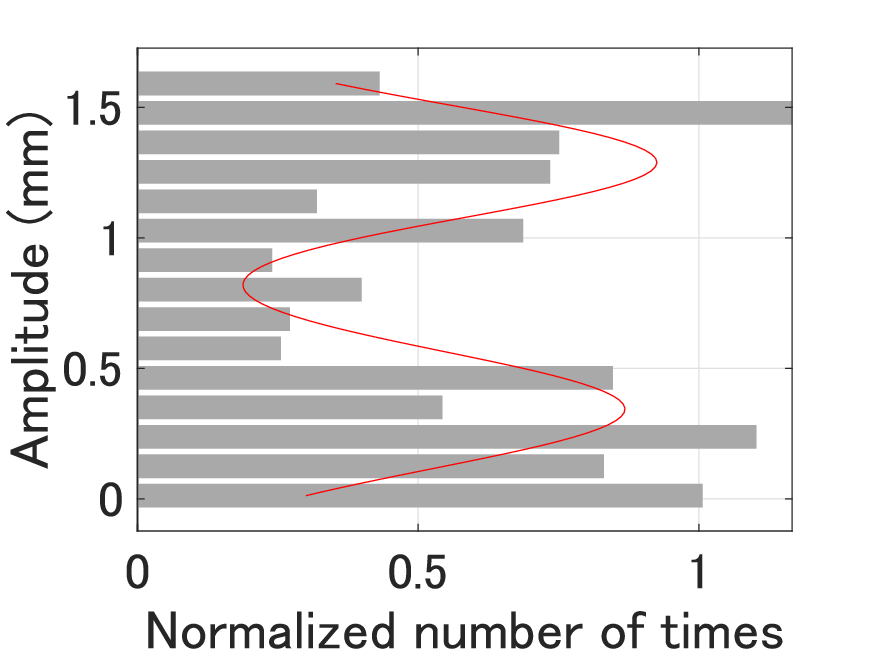}
                \\(b)
            \end{minipage}  
        \end{tabular}
        \\\vspace{2mm}
        \begin{tabular}{cc}
            \begin{minipage}[t]{0.23\textwidth}
                \centering
                \includegraphics[width = \textwidth]{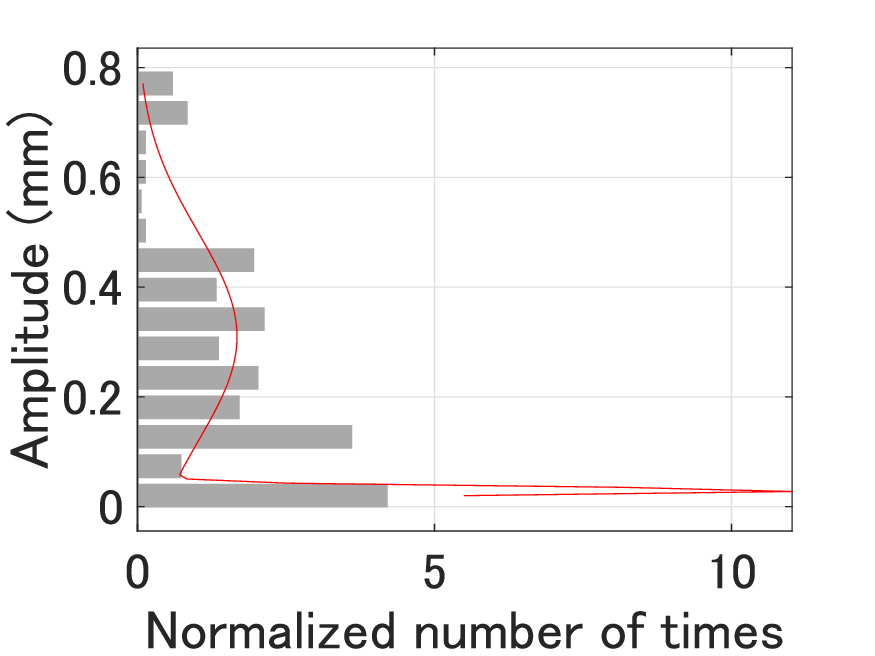}
                \\(c)
            \end{minipage}
            \begin{minipage}[t]{0.23\textwidth}
                \centering
                \includegraphics[width = \textwidth]{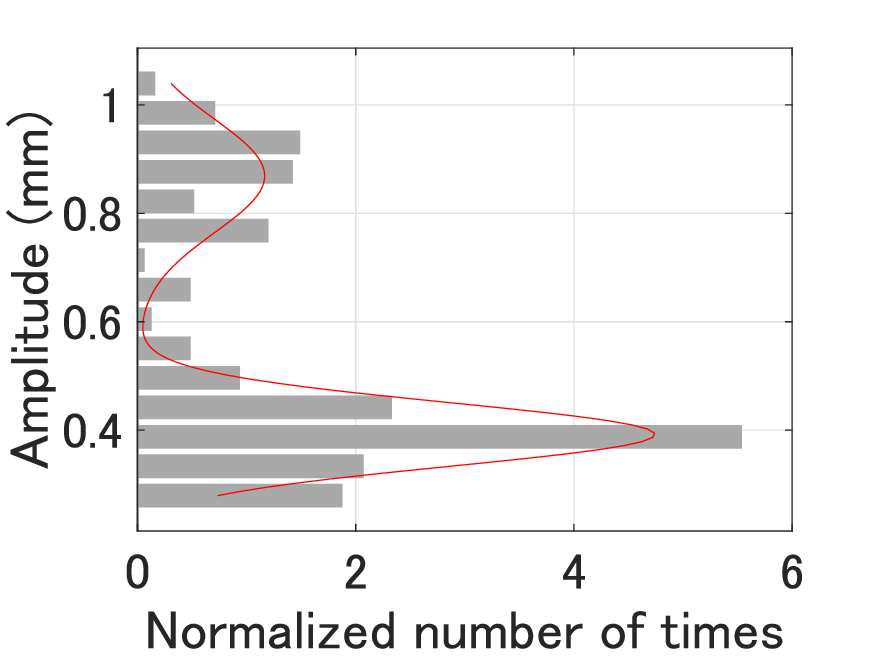}
                \\(d)
            \end{minipage}  
        \end{tabular}
    \end{center}
    \vspace{2mm}
    \caption{Probability density functions estimated using the EM algorithm for respiratory amplitudes when apnea and hypopnea events occur ((a) participant 1, (b) participant 3, (c) participant 4, and (d) participant 5).}
    \label{fig:EM_apnea}
\end{figure}

Fig. \ref{fig:EM_apnea} also shows example histograms of the respiratory amplitudes (gray bars) and the corresponding probability density functions when the apnea and hypopnea events occur, corresponding to Fig.~\ref{fig:EM_normal}. In this case, we can see two peaks in all panels and the peak gaps are larger than those observed in Fig.~\ref{fig:EM_normal} (i.e., $\mu_1/\mu_2$ is larger). The proposed method uses these characteristics to detect the apnea and hypopnea events automatically.

\subsection{Accuracy Evaluation of the Proposed Method}
This section provides a performance evaluation of the proposed method by comparing it with the conventional ABM. For this performance evaluation, we use the apnea-hypopnea index (AHI), which represents the combined number of apneas and hypopneas that occur per hour during sleep and is used to diagnose the severity of a patient's SAS. In addition, we also evaluate the accuracy of estimation of the number of apnea and hypopnea events per 30 min, which we then convert into the number per hour by simply doubling the number.

\begin{figure}[t]
    \centering
    \includegraphics[width=0.45\textwidth]{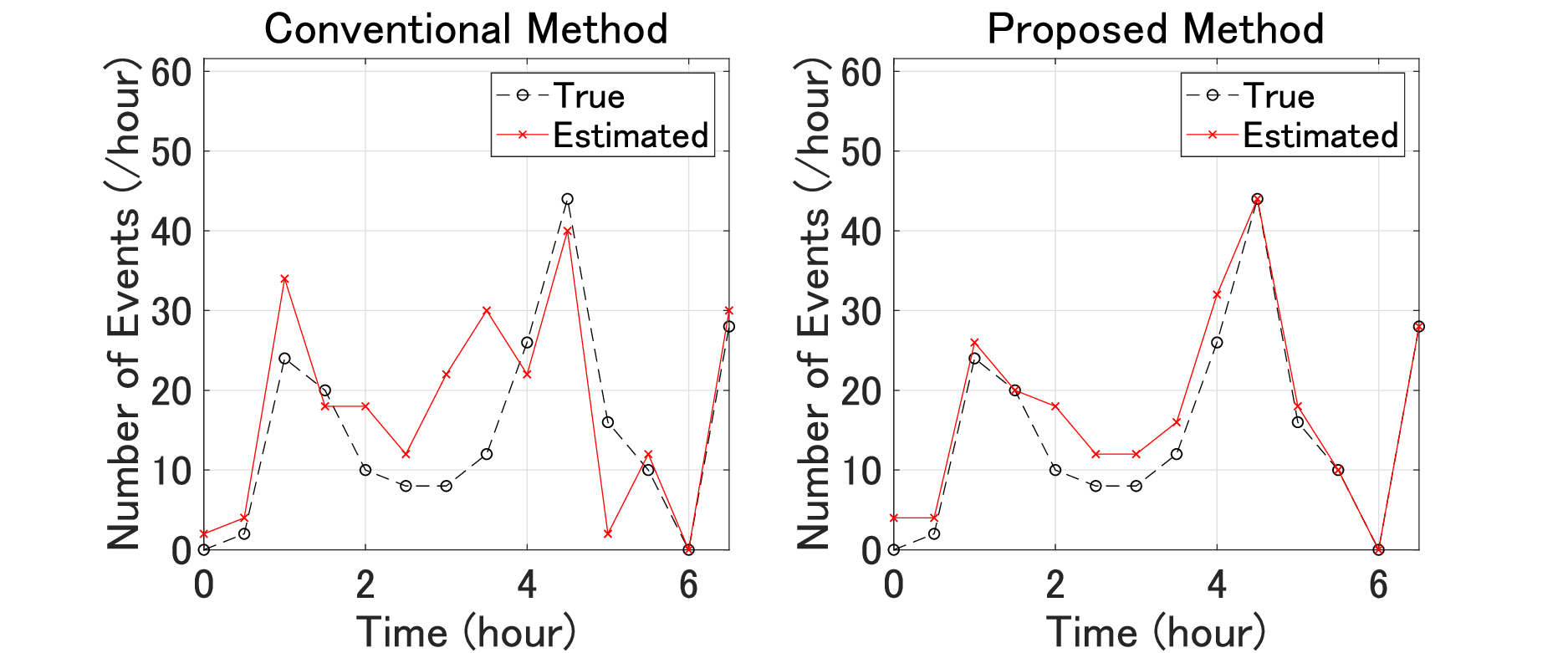}
    \\(a)\hspace{0.175\textwidth}(b)
    \caption{Number of apnea and hypopnea events estimated at 30 min intervals for participant 1 when using the (a) conventional and (b) proposed methods. }
    \label{fig:number_of_apnea1}
\end{figure}
\begin{figure}[t]
    \centering
    \includegraphics[width=0.45\textwidth]{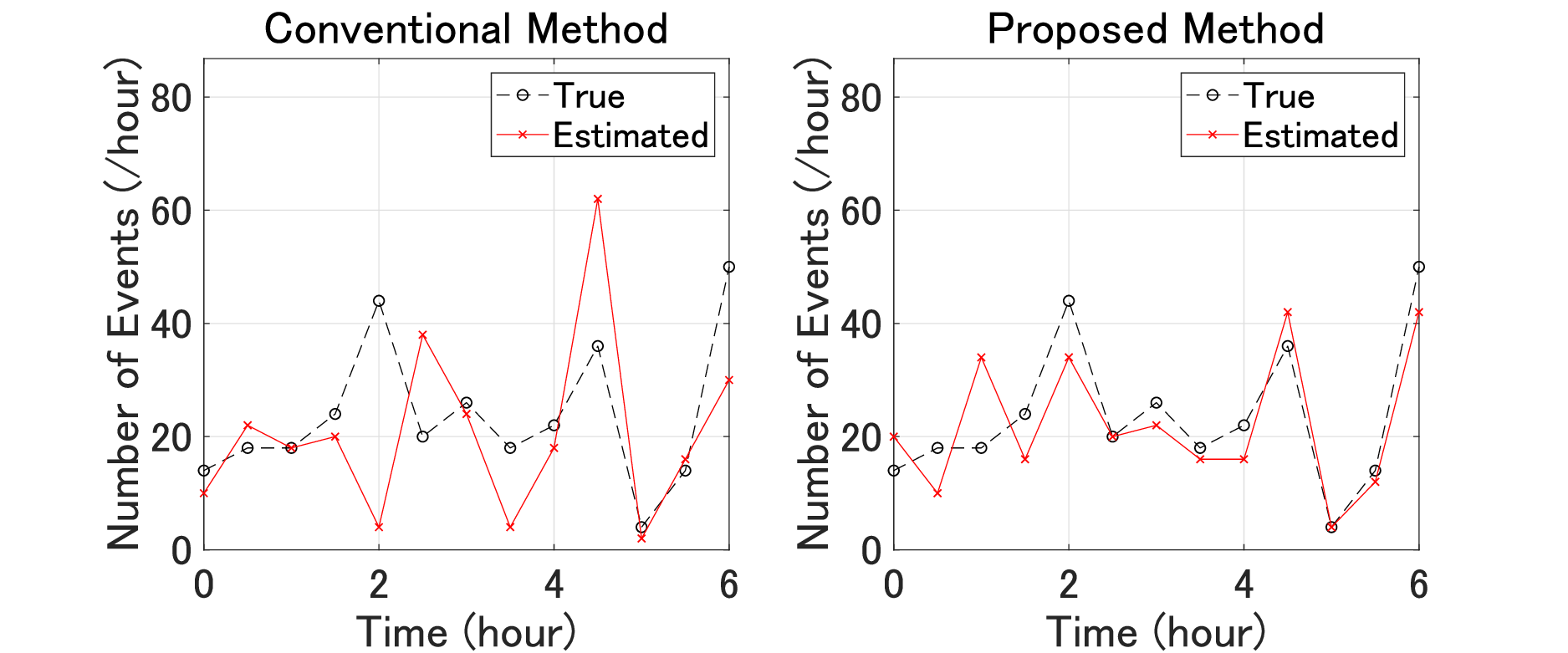}
    \\(a)\hspace{0.175\textwidth}(b)
    \caption{Number of apnea and hypopnea events estimated at 30 min intervals for participant 2 when using the (a) conventional and (b) proposed methods. }
    \label{fig:number_of_apnea2}
\end{figure}
\begin{figure}[t]
    \centering
    \includegraphics[width=0.45\textwidth]{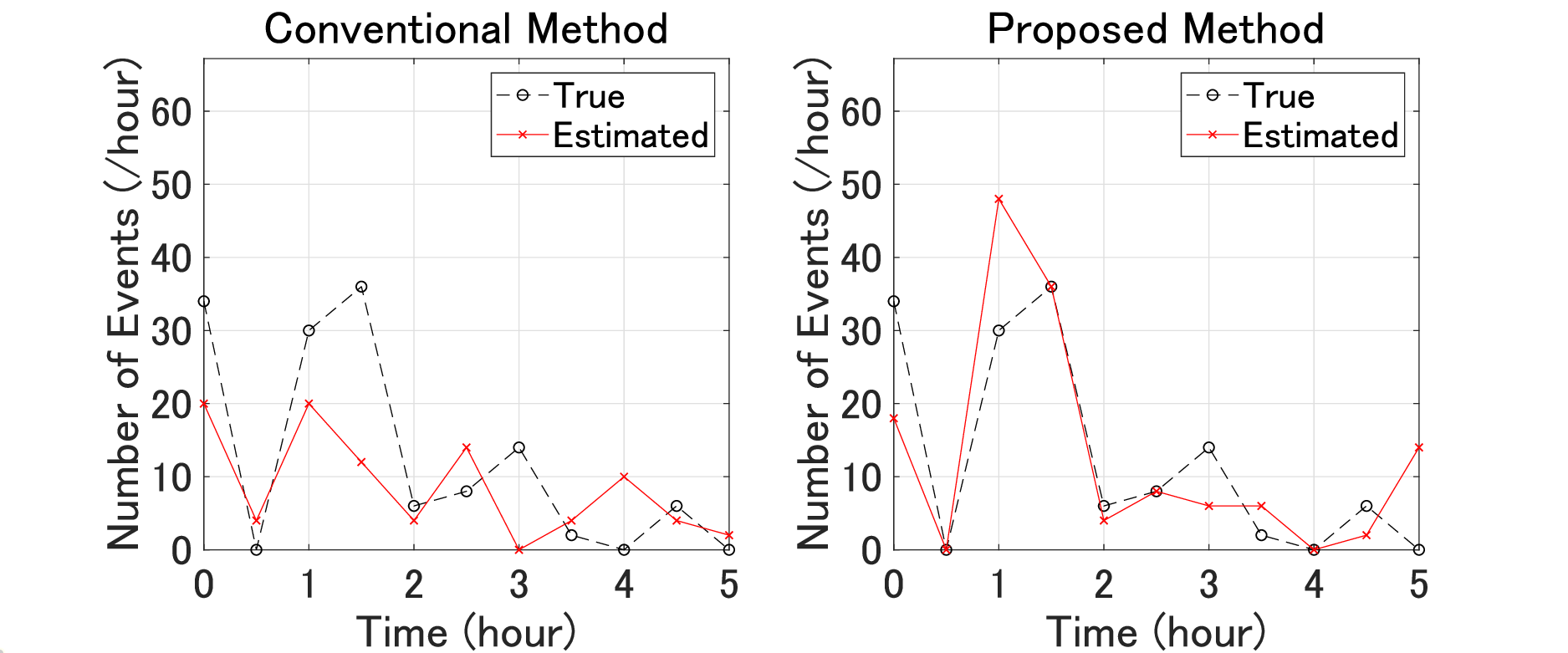}
    \\(a)\hspace{0.175\textwidth}(b)
    \caption{Number of apnea and hypopnea events estimated at 30 min intervals for participant 3 when using the (a) conventional and (b) proposed methods. }
    \label{fig:number_of_apnea3}
\end{figure}
\begin{figure}[t]
    \centering
    \includegraphics[width=0.45\textwidth]{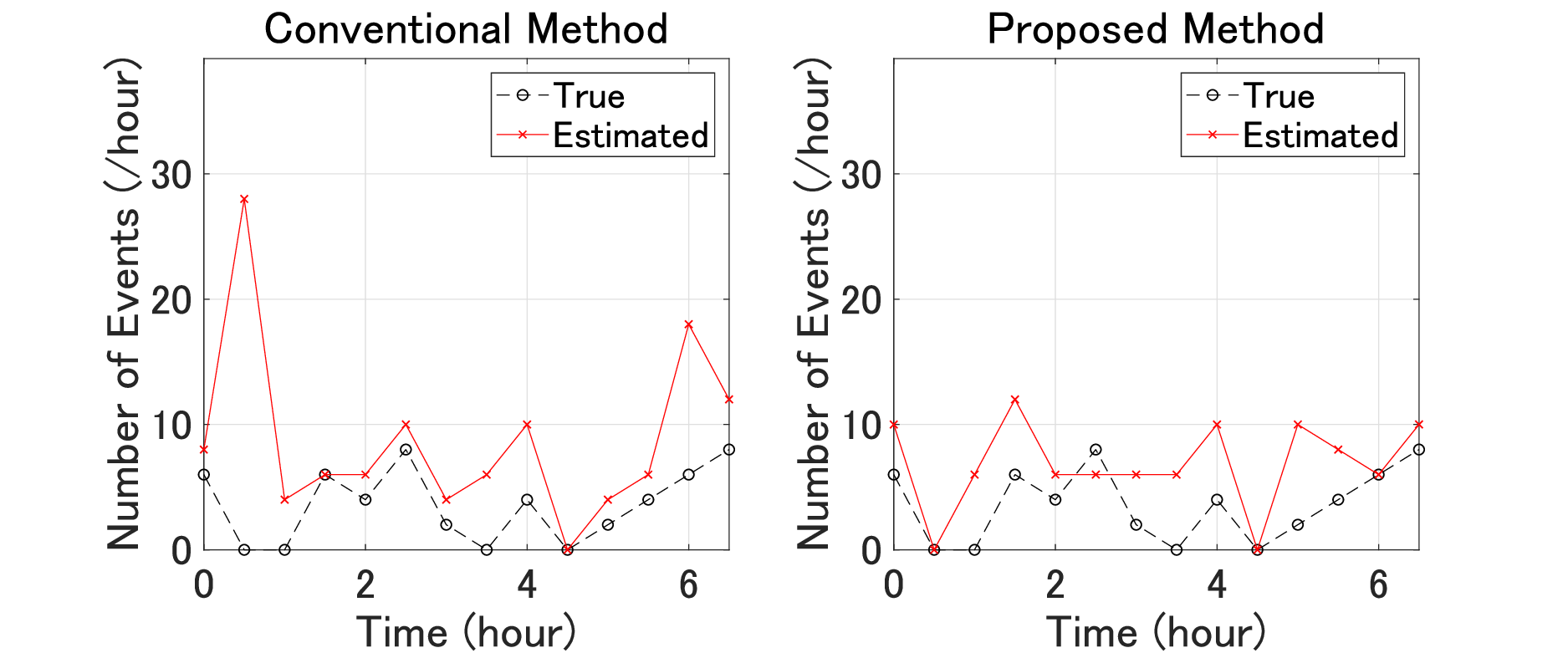}
    \\(a)\hspace{0.175\textwidth}(b)
    \caption{Number of apnea and hypopnea events estimated at 30 min intervals for participant 4 using the (a) conventional and (b) proposed methods. }
    \label{fig:number_of_apnea4}
\end{figure}
\begin{figure}[t]
    \centering
    \includegraphics[width=0.45\textwidth]{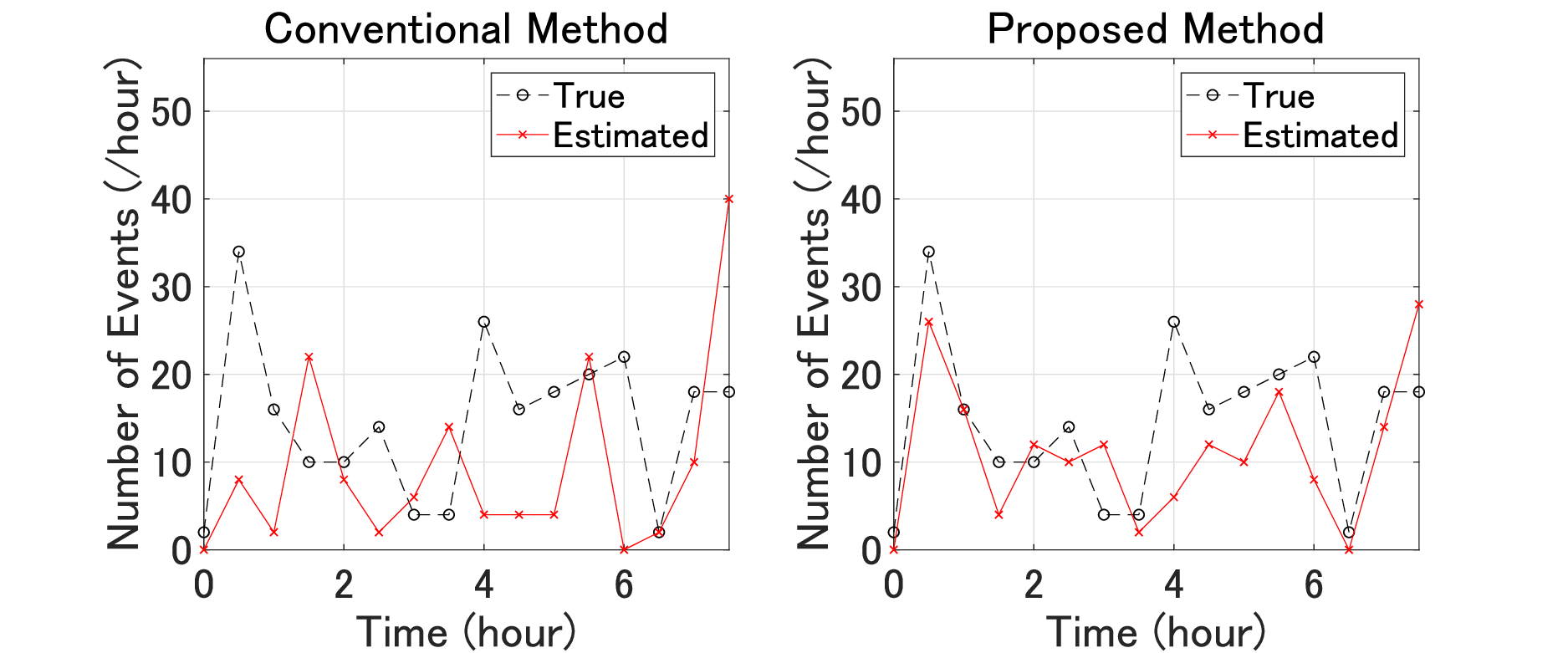}
    \\(a)\hspace{0.175\textwidth}(b) (b)
    \caption{Number of apnea and hypopnea events estimated at 30 min intervals for participant 5 using the (a) conventional and (b) proposed methods. }
    \label{fig:number_of_apnea5}
\end{figure}

Fig. \ref{fig:number_of_apnea1}--\ref{fig:number_of_apnea5} shows the numbers of apnea and hypopnea events per hour for each participant. In each figure, the red lines in panels (a) and (b) indicate the estimates from the conventional and proposed methods, respectively. The true numbers of apnea and hypopnea events, which are plotted as black dashed lines, were obtained from the PSG data and the doctor's diagnosis.

\begin{table}[t]
    \begin{center}
      \caption{Root-Mean-Square Errors of Estimated Numbers of Apnea and Hypopnea Events per Hour}
      \begin{tabular}{|c||c|c|} \hline
        Patient  & Conventional & Proposed\\
        number & method &method\\
        \hline
        1 & 8.2 & 3.5\\
        2 & 15.9 & 7.2\\
        3 & 10.6 & 8.9\\
        4 & 8.7 & 4.4\\
        5 & 14.0 & 7.9\\
        \hline
        Mean & 11.5 & 6.4\\
        
         \hline 
         
      \end{tabular}
      \label{tab:AHI}
    \end{center}
\end{table}
\begin{table}[t]
    \begin{center}
      \caption{AHI Estimated using the Conventional and Proposed Methods (Times/Hour).}
      \begin{tabular}{|c|c||c|c|c|c|} \hline
        \multicolumn{2}{|c|}{Patient} & \multicolumn{2}{c|}{Conventional method} & \multicolumn{2}{c|}{Proposed method}\\\hline
        No.&AHI&Estimated AHI&Error&Estimated AHI&Error\\
        \hline
        1& 30.4 & 35.9 & 5.5 & 35.6 & 5.2 \\
        2& 51.1 & 44.5 & 6.6 & 47.8 & 3.3 \\
        3& 27.2 & 18.8 & 8.4 & 28.4 & 1.2\\
        4& 7.4 & 18.2 & 10.8 & 14.2 & 6.8 \\
        5& 30.3 & 19.1 & 11.2 & 23.0 & 7.3 \\\hline
        Mean &-& - & 8.5 & - & 4.8\\
        
        \hline 
         
      \end{tabular}
      \label{tab:AHI1_2}
    \end{center}
\end{table}

Table \ref{tab:AHI} shows the root-mean-square (RMS) errors calculated for the estimated numbers of apnea and hypopnea events as shown in Figs. \ref{fig:number_of_apnea1}--\ref{fig:number_of_apnea5}. For all patients, the RMS error of the proposed method was shown to be smaller than that of the conventional method. The RMS error averaged over all patients was 11.5 times/hour when using the conventional method, whereas the corresponding value when using the proposed method was 6.4 times/hour, representing an improvement in accuracy of 1.8 times. 

Table \ref{tab:AHI1_2} summarizes the AHI values that were estimated using both the conventional and proposed methods. The AHI was estimated to have an average error of 8.5 times/hour when using the conventional method and 4.8 times/hour when using the proposed method, again resulting in an improvement of 1.8 times. Because the diagnostic criterion for sleep apnea is AHI$\geq$5 times/hour, the proposed method estimates the AHI with an accuracy of 4.8 times/hour, which reached the minimum accuracy requirement for diagnosis. These results illustrate the effectiveness of the proposed method for detection of apnea and hypopnea events in a noncontact manner using a radar system.

\section{Conclusion}
In this study, we have proposed a novel method for noncontact radar-based SA detection. The proposed method achieves adaptive apnea detection without prior knowledge of the respiratory amplitude of normal breathing. The proposed method uses a histogram of the respiratory amplitude as measured using a radar system, and then detects single or double peaks using the EM algorithm to distinguish the apnea and hypopnea events from normal breathing. To evaluate the performance of our proposed method, we performed experiments using both radar and PSG systems on five patients with SA symptoms. The experimental results showed that the proposed method estimated the AHI with an accuracy that was 1.8 times higher than that of the conventional method, confirming the effectiveness of the proposed method. In addition, we also demonstrated that the proposed method was able to estimate the AHI with a mean error of only 4.8 times/hour, which met the minimum accuracy requirement for use in SAS diagnosis. In conclusion, the proposed method was shown to be effective in using radar to perform noncontact apnea detection.

\begin{IEEEbiography}[{\includegraphics[height=1.25in,clip,keepaspectratio]{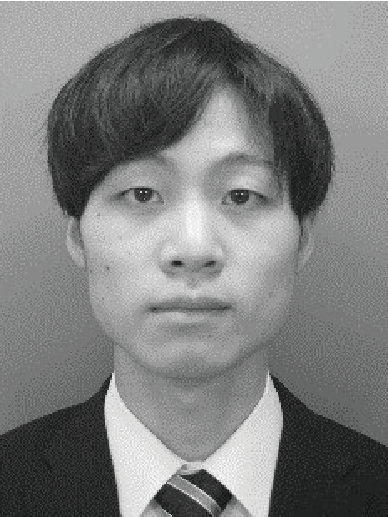}}]{Takato Koda}
  received a B.E.~degree in electrical and electronic engineering from Kyoto University, Kyoto, Japan, in 2020. He is currently working toward an M.E. degree in electrical engineering at the Graduate School of Engineering, Kyoto University.
\end{IEEEbiography}

\begin{IEEEbiography}[{\includegraphics[height=1.25in,clip,keepaspectratio]{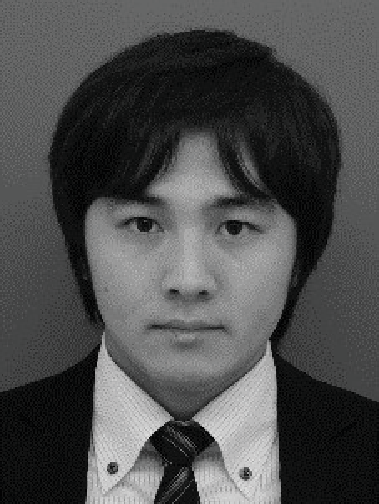}}]{Shigeaki Okumura}
    received his B.E. degree in electrical engineering from Kyoto University, Kyoto, Japan, in 2013 and M.I. and Ph.D. degrees in communications and computer engineering from the Graduate School of Informatics, Kyoto University, in 2015 and 2018, respectively. He has been with MaRI Co., Ltd. since 2019. His research interests include radar and audio signal processing and the noncontact measurement of vital signs.
  \end{IEEEbiography}
  
  \begin{IEEEbiography}[{\includegraphics[height=1.25in,clip,keepaspectratio]{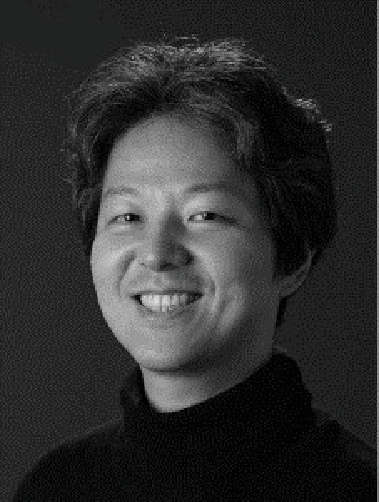}}]{Hirofumi Taki}
     (Member, IEEE) received the M.D. and Ph.D. degrees in informatics from Kyoto University, Japan, in 2000 and 2007, respectively. He was an Assistant Professor with the Graduate School of Informatics, Kyoto University, and an Associate Professor with the Graduate School of Biomedical Engineering, Tohoku University. He founded MaRI Co., Ltd. in 2017, and has been the CEO ever since. His research interests include digital signal processing in measurement of biological information.
  \end{IEEEbiography}

  \begin{IEEEbiography}[{\includegraphics[height=1.25in,clip,keepaspectratio]{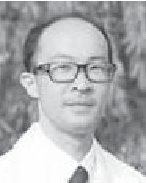}}]{Satoshi Hamada} received Ph.D. degree in respiratory medicine from Kyoto University Graduate School of Medicine in 2016. He is currently an Assistant Professor of the Department of Advanced Medicine for Respiratory Failure, Graduate School of Medicine, Kyoto University Hospital. 
  \end{IEEEbiography}

  \begin{IEEEbiography}[{\includegraphics[height=1.25in,clip,keepaspectratio]{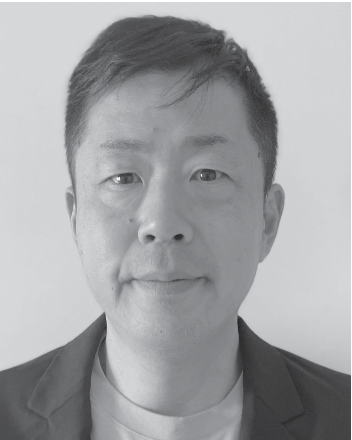}}]{Hironobu Sunadome} is currently an Assistant Professor of the Department of Respiratory Care and Sleep Control Medicine, Graduate School of Medicine, Kyoto University. His sub-specialty is respiratory medicine, allergic medicine, and sleep breathing medicine. He received his Ph.D. (Respiratory medicine) in 2021.
  \end{IEEEbiography}
  
  \begin{IEEEbiography}[{\includegraphics[height=1.25in,clip,keepaspectratio]{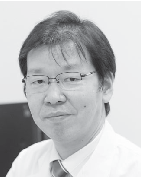}}]{Susumu Sato} is currently an Associate Professor of the Department of Respiratory Care and Sleep Control Medicine, Graduate School of Medicine, Kyoto University. He is a head of a section of the Sleep Laboratory of Kyoto University Hospital. His sub-specialty is respiratory medicine, and he received his Ph.D. in 2004. His current research fields are respiratory medicine, respiratory mechanics, biomedical engineering, and image analysis, including Outcome research. He is a member of ATS, ERS, APSR, AARC, JRS, JSSR, JSRCM, JSRCR, JARM, JSA, and JSIM. He is a deputy head of the Assembly of Respiratory Structure and Function of APSR (Asia-Pacific Society of Respirology).
 \end{IEEEbiography}

 \begin{IEEEbiography}[{\includegraphics[height=1.25in,clip,keepaspectratio]{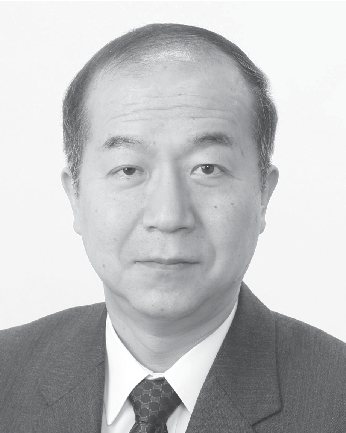}}]{Kazuo Chin} graduated from the Kyoto University School of Medicine in 1981 and received his doctoral degree in internal medicine from Kyoto University Graduate School of Medicine in 1990. He was a program Specific Professor in the Department of Respiratory Care and Sleep Control Medicine, Graduate School of Medicine, Kyoto University from 2008 to 2021. He is currently a Professor in the Department of Sleep Medicine and Respiratory Care, Division of Sleep Medicine, Nihon University of Medicine. At the same time, he also works part-time in the Department of Human Disease Genomics, Center for Genomic Medicine, Graduate School of Medicine, Kyoto University. His research interests are respiratory care and sleep medicine, including sleep disordered breathing.
 \end{IEEEbiography}

\begin{IEEEbiography}[{\includegraphics[height=1.25in,clip,keepaspectratio]{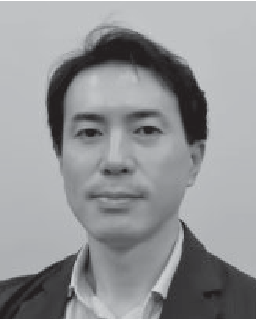}}]{Takuya Sakamoto} (Senior Member, IEEE) received a B.E. degree in electrical and electronic engineering from Kyoto University, Kyoto, Japan, in 2000 and M.I. and Ph.D. degrees in communications and computer engineering from the Graduate School of Informatics, Kyoto University, in 2002 and 2005, respectively.
  
From 2006 through 2015, he was an Assistant Professor at the Graduate School of Informatics, Kyoto University. From 2011 through 2013, he was also a Visiting Researcher at Delft University of Technology, Delft, the Netherlands. From 2015 until 2019, he was an Associate Professor at the Graduate School of Engineering, University of Hyogo, Himeji, Japan. In 2017, he was also a Visiting Scholar at University of Hawaii at Manoa, Honolulu, HI, USA. From 2019 until 2022, he was an Associate Professor at the Graduate School of Engineering, Kyoto University. From 2018 through 2022, he was also a PRESTO researcher of the Japan Science and Technology Agency, Japan. From 2022, he has been a Professor at the Graduate School of Engineering, Kyoto University. His current research interests lie in wireless human sensing, radar signal processing, and radar measurement of physiological signals.

Prof. Sakamoto was a recipient of the Best Paper Award from the International Symposium on Antennas and Propagation (ISAP) in 2004, the Young Researcher's Award from the Information and Communication Engineers of Japan (IEICE) in 2007, the Best Presentation Award from the Institute of Electrical Engineers of Japan in 2007, the Best Paper Award from the ISAP in 2012, the Achievement Award from the IEICE Communications Society in 2015, 2018, and 2023, the Achievement Award from the IEICE Electronics Society in 2019, the Masao Horiba Award in 2016, the Best Presentation Award from the IEICE Technical Committee on Electronics Simulation Technology in 2022, the Telecom System Technology Award from the Telecommunications Advancement Foundation in 2022, and the Best Paper Award from the IEICE Communication Society in 2007 and 2023. 
\end{IEEEbiography}

\end{document}